\documentclass[12pt]{iopart}
\usepackage{graphicx}
\usepackage{booktabs}
\usepackage{iopams}
\usepackage[pdfusetitle,
bookmarks=true,colorlinks=true, linkcolor=blue, citecolor=blue, linktoc=page]{hyperref}
\usepackage{comment}
\usepackage{mhchem}
\usepackage[dvipsnames]{xcolor}
\usepackage{subcaption}
\usepackage[shortlabels]{enumitem}
\usepackage{multirow}
\usepackage{multicol}

\begin{document}
\title[]{Impact of triangularity on edge transport and divertor detachment: a SOLPS-ITER study of TCV L-mode plasmas}

\author{F. Mombelli$^{1}$, A. Mastrogirolamo$^{1}$, E. Tonello$^{2}$, O. Février$^{2}$, G. Durr-Legoupil-Nicoud$^{2}$, M. Carpita$^{2}$, F. Subba$^{3}$, M. Passoni$^{1}$, the TCV team$^{4}$ and the EUROfusion Tokamak Exploitation Team$^{5}$}

\address{$^{1}$ Politecnico di Milano, Department of Energy, Milan, 20133, Italy}
\address{$^{2}$ Ecole Polytechnique Fédérale de Lausanne, Swiss Plasma Center, Lausanne, 1015, Switzerland}
\address{$^{3}$ NEMO Group, Dipartimento Energia, Politecnico di Torino, Turin, 10129, Italy}
\address{$^{4}$ See the author list of Duval et al 2024 Nucl. Fusion 64 112023}
\address{$^{5}$ See the author list of Joffrin et al 2024 Nucl. Fusion 64 112019}

\ead{fabio.mombelli@polimi.it}

\providecommand{\keywords}[1]
{
  \small	
  \textbf{\textit{Keywords---}} #1
}

\begin{abstract}
Negative triangularity (NT) magnetic configurations have recently gained attention as a promising route to achieve H-mode-like confinement without edge-localized modes (ELMs) and without a power threshold for access. While both core and edge confinement properties of NT have been extensively documented, consistently lower divertor target cooling and increased difficulty in achieving a detached regime have been observed. This work presents a comparative SOLPS-ITER modeling study of two Ohmic L-mode discharges in the TCV tokamak with identical divertor geometry and opposite upper triangularity. We investigate whether magnetic geometry alone can account for the experimentally observed differences in plasma detachment behavior. Simulations with identical transport coefficients reveal no significant differences between NT and positive triangularity (PT) cases, even when including drifts. A parametric scan of radial anomalous transport coefficients shows that reproducing the experimental profiles requires lower particle diffusivity in NT, consistent with reduced turbulent transport and previous findings. Furthermore, the evolution of simulated neutral pressures and recycling fluxes along a density scan reproduces experimental observations of larger neutral divertor pressure in PT, highlighting a distinct neutral dynamics in the two cases. These results support the interpretation that altered cross-field transport, rather than magnetic geometry alone, underlies the observed differences in divertor behavior between NT and PT scenarios.
\end{abstract}


\section{Introduction}

In the pursuit of sustainable energy via magnetic confinement fusion, the high-confinement mode (H-mode) has long been considered as the most promising regime for tokamak operation due to its superior plasma confinement capabilities \cite{Wagner1982, 1989}. This improved confinement arises from the formation of steep edge pressure gradients across the separatrix, which trigger Edge Localized Modes (ELMs)—bursty magneto-hydrodynamic (MHD) instabilities that expel particles and energy into the scrape-off layer (SOL) \cite{Hill1997, leonard_edge-localized-modes_2014}. While ELMs help prevent impurity accumulation \cite{leonard_edge-localized-modes_2014}, the associated heat fluxes can severely damage divertor components \cite{Evans2013}, making their mitigation or suppression a key objective for the next generation of fusion reactors.

Among the most promising ELM-free alternatives are magnetic configurations with negative triangularity (NT), characterized by an inward-pointing D shape \cite{marinoni_brief_2021}, as opposed to the positive triangularity (PT) configurations adopted in standard scenarios such as ITER’s baseline \cite{Labit2024}. Experimental campaigns on TCV, DIII-D, and ASDEX Upgrade \cite{Pochelon1999, Camenen2007, coda_enhanced_2022, austin_achievement_2019, marinoni_diverted_2021, happel_overview_2023} have shown that NT plasmas can achieve H-mode-like confinement while operating in the inherently ELM-free L-mode \cite{fontana_effects_2020}.

Most studies on NT have focused on core confinement, highlighting turbulence suppression as the main driver for performance improvements \cite{Camenen2007, Merlo2015, Giannatale2022}. Yet, assessing NT’s potential as a reactor-relevant configuration also requires understanding edge plasma behavior, particularly in relation to power exhaust and access to detachment \cite{leonard_plasma_2018}. In this direction, turbulent fluid simulations have shown that NT shaping stabilizes turbulence in the SOL \cite{Lim2023, Riva2017}, aligning with experimental observations indicating a suppression of first wall interaction for sufficiently negative triangularity values \cite{Han2021}.

A recent study by Février et al. \cite{Fvrier2024} experimentally investigated detachment access in L-mode plasma scenarios across a wide range of upper and lower triangularities in the TCV tokamak. The study found that, in all cases, NT shaping resulted in more challenging detachment access and reduced divertor cooling compared to PT configurations \cite{Fvrier2024}. This is consistent with the observed reduction in the power fall-off length (\(\lambda_q\)) in NT compared to PT scenarios, which may lead to more localized and concentrated heat fluxes at the divertor targets \cite{Scotti2024,faitsch_dependence_2018,Laribi2021}.  

On the front of numerical analysis conducted with mean-field codes, Muscente et al. \cite{muscente_analysis_2023} showed using SOLEDGE2D-EIRENE that increasingly negative upper triangularity in TCV discharges required a monotonic reduction of core particle diffusivity to match experimental profiles, while no clear trend emerged for heat diffusivity.

More recently, the work by Tonello et al. \cite{Tonello2024} investigated two TCV L-mode discharges differing both in triangularity and divertor geometry using the SOLPS-ITER code. By employing fixed transport coefficients, the study aimed at isolating the impact of magnetic geometry on plasma profiles, revealing differences in transport and neutral particle accumulation in the scrape-off layer. These findings reproduced the experimental observation of a hotter and more attached outer target in the NT scenario compared to its PT counterpart \cite{Tonello2024}.

As a direct follow-up to the work of Tonello et al. \cite{Tonello2024}, the present study advances the numerical investigation of transport properties and detachment behavior in NT versus PT scenarios, aiming to address one of the key open questions identified in that paper. To this end, two Ohmic L-mode discharges in TCV, characterized by identical magnetic geometry in the divertor region but opposite upper triangularity, are analyzed using the SOLPS-ITER code. These discharges, also part of the experimental dataset discussed in \cite{Fvrier2024}, offer the opportunity to isolate the effect of triangularity from that of divertor geometry—which, unlike in \cite{Tonello2024}, is approximately the same.

More specifically, the aim of this paper is to assess, using the SOLPS-ITER code, whether the sole difference in upper triangularity can explain the experimentally observed variations in plasma parameter profiles—similarly to what was achieved in \cite{Tonello2024}—or whether ad hoc assumptions on the anomalous transport regime must be introduced to describe and interpret the experimental evidence.

The paper is organized as follows: Section \ref{reference_exp_frame} presents an overview of the experimental features of the TCV discharges considered in this study. Section \ref{sim_setup} details the SOLPS-ITER simulation setup and modeling approach, with the corresponding results discussed in Section \ref{results}. Finally, Section \ref{conclusions} draws the main conclusions of the study. This article is also accompanied by the \ref{appendix} discussing technical notes on the use of drifts in the presented simulations.

\section{Reference experimental framework}
\label{reference_exp_frame}

\begin{figure} 
  \centering
  \includegraphics[width=\textwidth]{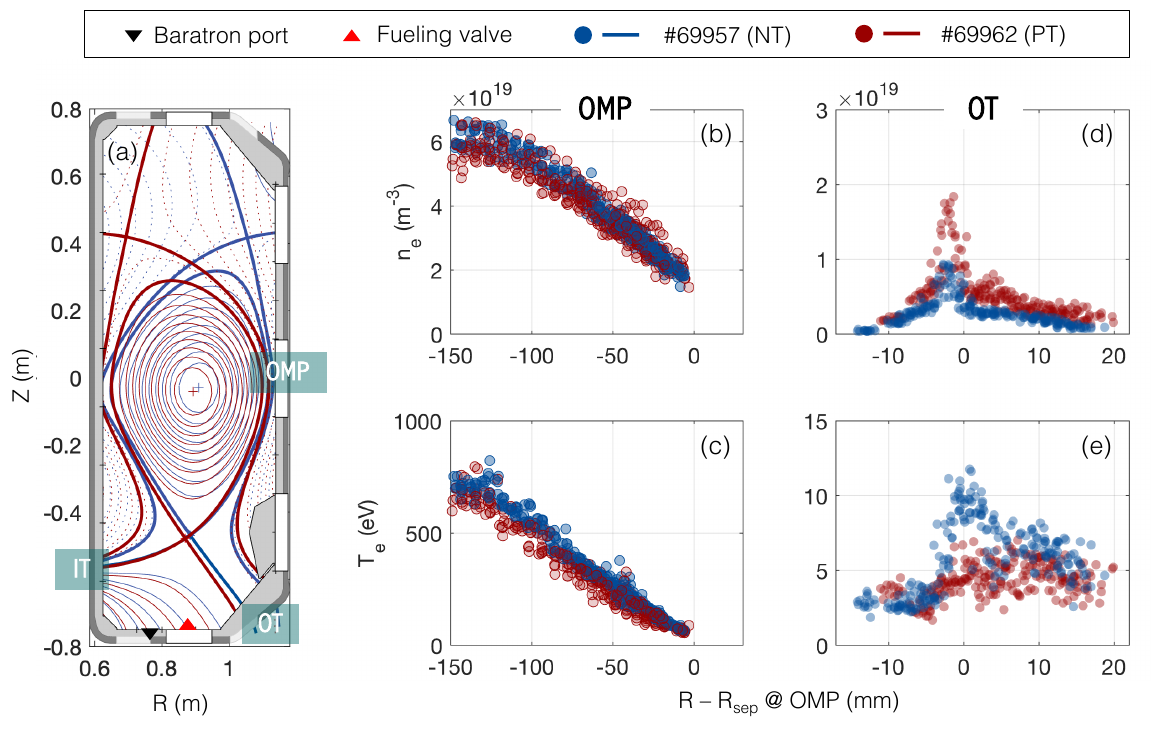}
  \caption{(a) Reconstruction of the magnetic equilibria for the discharges under investigation, which exhibit a superimposable divertor geometry and opposite upper triangularity. (b)-(c) Electron density and temperature data, respectively, as a function of the radial distance from the separatrix at the OMP for the two discharges, obtained via Thomson Scattering within the reference time windows, i.e. [1.18 s, 1.28 s] for NT and [1.10 s, 1.20 s] for PT. (d)-(e) Electron density and temperature data, respectively, at the OT, acquired using Langmuir probes within the reference time windows and represented as a function of the distance from the separatrix mapped at the OMP.}
  \label{fig:mag_eq_exp_data}
\end{figure}

In this section, the key experimental characteristics of the TCV discharges considered in this study - which are more extensively described in Reference \cite{Fvrier2024} - are briefly reviewed. The pulses analyzed are Ohmic L-mode deuterium-only discharges \#69957 and \#69962, with their magnetic equilibria at selected time instants shown in Figure \ref{fig:mag_eq_exp_data}(a). These two discharges share the same divertor geometry and have a common lower triangularity close to zero ($\delta_{bot} = -0.02$), while the upper triangularity is either negative ($\delta_{top} = -0.3$) or positive ($\delta_{top} = 0.2$), respectively. In both cases, the ohmic heating power is $P_{\text{ohm}} \simeq 230$ kW and the toroidal magnetic field is in the forward direction, i.e. the one for which the $\vec{B} \times \nabla B$ drift for ions points from the core towards the X-point. The parallel connection lengths from the outer midplane (OMP) to the outer target (OT) are very similar in the two cases ($L_{\parallel} = 14.7$ m in NT and $L_{\parallel} = 15.7$ m in PT).

Both discharges investigate divertor detachment through a density ramp in which the magnetic equilibrium remains constant. As the average electron density \(\langle n_e \rangle\) increases during the discharge, the peak electron temperature \(T_e\) measured at the OT decreases in both cases. However, while in the PT case the target temperature drops below 5 eV—indicating the onset of detachment—it remains above 5 eV in the NT case, suggesting that the plasma remains attached throughout the entire discharge.

Figure \ref{fig:mag_eq_exp_data}(b)-(e) shows the plasma profiles collected within a 0.1 s time window centered around two \textit{reference} time points, one for each discharge (\(t_{\text{ref}}^{\text{NT}} = 1.23\) s, \(t_{\text{ref}}^{\text{PT}} = 1.15\) s), selected so that the average densities \(\langle n_e \rangle\) in both cases are essentially the same. The upstream electron density and temperature data are obtained via Thomson scattering (TS). However, for the specific magnetic equilibria under consideration, the TS line of sight grazes the X-point, thereby restricting the radial extent of the experimental data to the range \(R - R_{\text{sep}} < 0\). Nonetheless, the upstream profiles within this region appear to be well-aligned and largely overlapping. In contrast, the profiles at the OT measured via Langmuir probes clearly show a higher \(n_e\) and lower \(T_e\) in the PT case compared to NT, consistent with the observations made above. 

\begin{figure} 
  \centering
  \includegraphics[width=0.45\textwidth]{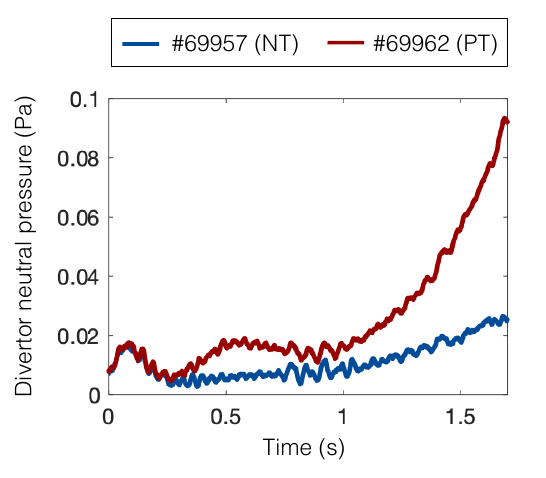}
  \caption{Experimental evolution of neutral pressure measured at the divertor baratron gauge during the density ramp for the discharges under investigation.}
  \label{fig:exp_neut_press}
\end{figure}

An additional experimental aspect of interest for this study is the evolution of neutral species pressure at the divertor during the density ramp, measured by TCV's divertor baratron gauge. While both discharges show a pressure increase with rising average plasma density, the measurements (Figure \ref{fig:exp_neut_press}) indicate consistently higher neutral pressure in the PT case compared to NT.

\section{Simulation setup and strategy}
\label{sim_setup}
This work utilizes the SOLPS-ITER code (v. 3.0.9), a state-of-the-art tool for modeling the boundary plasma in magnetic fusion devices \cite{wiesen_new_2015, BONNIN2016}. The package combines two main modules: B2.5 \cite{Rozhansky2001, Rozhansky2009}, a multi-fluid, non-turbulent code that solves particle, momentum, and energy conservation of charged plasma species under the assumption of toroidal symmetry; and EIRENE \cite{reiter_eirene_2005}, a kinetic Monte Carlo code that simulates the transport of neutral species, which are invariably present in the boundary plasma.

\begin{figure} 
  \centering
  \includegraphics[width=0.55\textwidth]{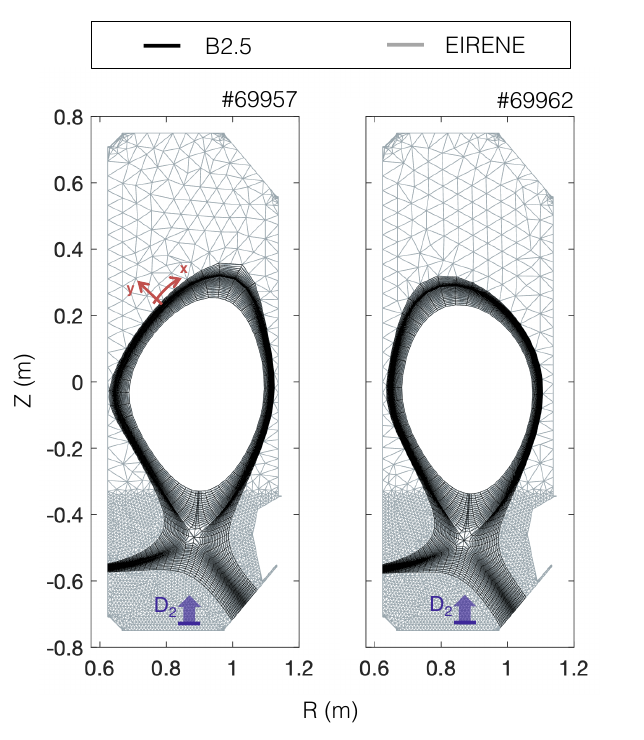}
  \caption{Computational meshes of the B2.5 (structured quadrilateral) and EIRENE (unstructured triangular) codes for the two discharges under investigation, constructed based on the reconstruction of the respective magnetic equilibria displayed in Figure \ref{fig:mag_eq_exp_data}(a). The location of the \ce{D2} gas puff is indicated.}
  \label{fig:mesh_comp}
\end{figure}

Figure \ref{fig:mesh_comp} displays the simulation domains, which comprise a 2D poloidal cross-section of the tokamak, bounded by a realistic representation of the vessel structures. The B2.5 computational mesh is a structured quadrangular grid aligned with the magnetic flux surface, as reconstructed from the magnetic equilibrium. It extends poloidally between the two targets and radially from inside the separatrix to the innermost flux surface tangent to a non-divertor first wall structure. In contrast, the EIRENE mesh is an unstructured triangular grid that covers the B2.5 mesh and extends up to the device walls. Figure \ref{fig:mesh_comp} illustrates the B2.5 (black) and EIRENE (grey) computational meshes, based on the magnetic equilibria of the two oppositely triangulated discharges shown in Figure \ref{fig:mag_eq_exp_data}(a). The two B2.5 meshes exhibit a similar radial extent both at the targets and in the SOL, and share the same spatial resolution, i.e., \(n_x \times n_y = 84 \times 36\), where \(n_x\) and \(n_y\) are the poloidal and radial resolutions, respectively.

The simulations presented in this work include deuterium (D) and carbon (C) as plasma species. While deuterium constitutes the primary component of the plasma, the presence of carbon arises from impurities sputtered from the graphite walls. Specifically, the ionic and neutral species considered are \ce{D}, \ce{D+}, \ce{D2}, \ce{D2+}, neutral carbon \ce{C}, and its six ionization states \ce{C^n+}.

B2.5 solves a density and parallel momentum balance equation for each ion species, an electron energy balance equation, a single ion energy balance equation assuming a common ion temperature \(T_i\) and a continuity equation for electric current. Accordingly, the following set of boundary conditions—common to both NT and PT simulations—is applied along the boundaries of the B2.5 computational mesh:

\begin{itemize}
    \item \textit{Divertor targets}: sheath boundary conditions for densities, temperatures, velocities and electrostatic potential. Since perfect entrainment of heavier \ce{C} impurities is assumed, the sound velocity imposed at the targets is common to all ion species and weighted over $\sqrt{n_e / \sum_a n_a m_a}$, $n_a$ and $m_a$ being the density and mass of the $a$-th ion species.
    
    \item \textit{Far SOL boundary and Private Flux region (PFR)}: zero gradients of the parallel velocities, zero current for the potential equation and decay boundary condition for $n_a$, $T_e$ and ion temperature $T_i$ enforcing a characteristic radial decay length $\lambda_{\text{decay}} = 1$ cm. For cases with drifts, the decay boundary conditions are replaced by drift-compatible radial leakage boundary conditions, which set the radial loss of particles for all species, as well as electron and ion energy fluxes, to a small fraction \(\alpha = 10^{-3}\) of the corresponding fluxes to a solid surface.
    
    \item \textit{Core}: zero gradients of the parallel velocity, zero current for the potential equation. Energy fluxes for electrons and ions are imposed such that the total power flowing into the SOL from the core is $P_{ \text{tot}} = P_{\text{Ohm}} - P_{ \text{rad,core}} \simeq 180$ kW. It is also imposed that the flux of \ce{D+} ions from the core to the SOL is equal and opposite to the flux of neutrals reaching the core edge (ionising core), i.e. $\Gamma_{\text{core,} \ce{D+}} = - \left (2 \Gamma_{\text{core,} \ce{D2}} + \Gamma_{\text{core,} \ce{D}} \right )$. The total flux of \ce{C} ions is set to zero.
\end{itemize}

The pumping associated with the absorption capabilities of carbon walls is modeled by specifying a recycling (or albedo) coefficient \(R = 0.99\) for all species, consistent with previous TCV simulations \cite{Wensing2019, Wensing2021, Tonello2024}.

The physical and chemical sputtering of \ce{D+} species on \ce{C} walls are responsible for the release of \ce{C} impurities into the plasma. The physical sputtering yield depends on the energy and angle of the impinging projectile and is calculated using the Roth-Bohdansky formula \cite{Bohdansky1980}. For chemical sputtering, a constant yield of 3.5\% is applied, as in previous TCV simulations \cite{Wensing2019, Wensing2021, Tonello2024}.

The modeling of the \ce{D2} gas puff requires specifying a puffing surface on the EIRENE mesh, Figure \ref{fig:mesh_comp}, corresponding to the experimental position of the gas puff valve. Instead of manually controlling the influx of \ce{D2} molecules (\(\Gamma_{\ce{D2}}\)), a feedback scheme is implemented. In this approach, \(\Gamma_{\ce{D2}}\) is internally adjusted by the code to ensure that the plasma state converges to a prescribed value of electron density at the outer midplane separatrix (\(n_{e,\text{sep}}\)), specified for each simulation.

When drifts are turned on in the simulations, cross-field drifts ($E \times B$ and diamagnetic), as well as currents, are considered. Activating drifts in SOLPS-ITER requires careful handling due to the potential for fatal numerical instabilities, including the need for reduced time steps, the implementation of drift-compatible boundary condition schemes, and the adoption of convergence speed-up techniques to address otherwise prohibitively long computational times. Among the known drift convergence acceleration methods, detailed in Reference \cite{Kaveeva2018}, the partial flux surface averaging method has been used in our drift-enabled simulations. The discussion of additional technical aspects related to the activation of drifts in these simulations is deferred to \ref{appendix}.

Anomalous cross-field transport in SOLPS-ITER is treated using a diffusive mean-field approximation, which requires specifying diffusive transport coefficients in the simulation setup. These coefficients include particle diffusivity (\(D_n\)) and heat diffusivity for ions (\(\chi_i\)) and electrons (\(\chi_e\)), expressed in \( \text{m}^2 \, \text{s}^{-1} \).

The simulation strategy underlying this work is articulated in three successive steps. In continuity with the study by Tonello et al. \cite{Tonello2024}, initial simulations are performed using the two computational meshes—each corresponding to a different upper triangularity—while employing identical input parameters for the simulations, in particular the same anomalous transport coefficients. More advanced physics models (e.g., drifts, poloidally varying transport anomalous diffusivities) are progressively enabled during this phase.

Anticipating that identical transport coefficients will not be sufficient to reproduce the experimentally observed differences between NT and PT, a parametric scan of the radial transport coefficients is then carried out to identify a set of values capable of reconciling the simulations with the reference experimental conditions shown in Figure \ref{fig:mag_eq_exp_data}. These optimized transport coefficients are finally employed to simulate a density ramp via variation of \(n_{\text{e,sep}}\), allowing for the investigation of the evolution of neutral pressure in the divertor region.

\section{Results and discussion}
\label{results}

\subsection{Effect of magnetic geometry}
\label{magnetic_geom}

In line with the simulation strategy adopted, this subsection presents the results of a set of homologous simulations, carried out under identical input conditions but using the computational mesh specific to each discharge. The same values of \( n_{e,\text{sep}} \) and anomalous transport coefficients were applied, consistent with previous TCV modeling studies, namely \( D_n = 0.2 \) m\(^2\)/s and \( \chi_e = \chi_i = 1.0 \) m\(^2\)/s \cite{Wensing2021,Tonello2024}. Since the shape of the SOL profiles is influenced by the position at which the far-SOL boundary conditions are applied, it is worth stressing that both computational meshes were constructed with the same radial extent to ensure a consistent and meaningful comparison.

\begin{figure} 
  \centering
  \includegraphics[width=0.9\textwidth]{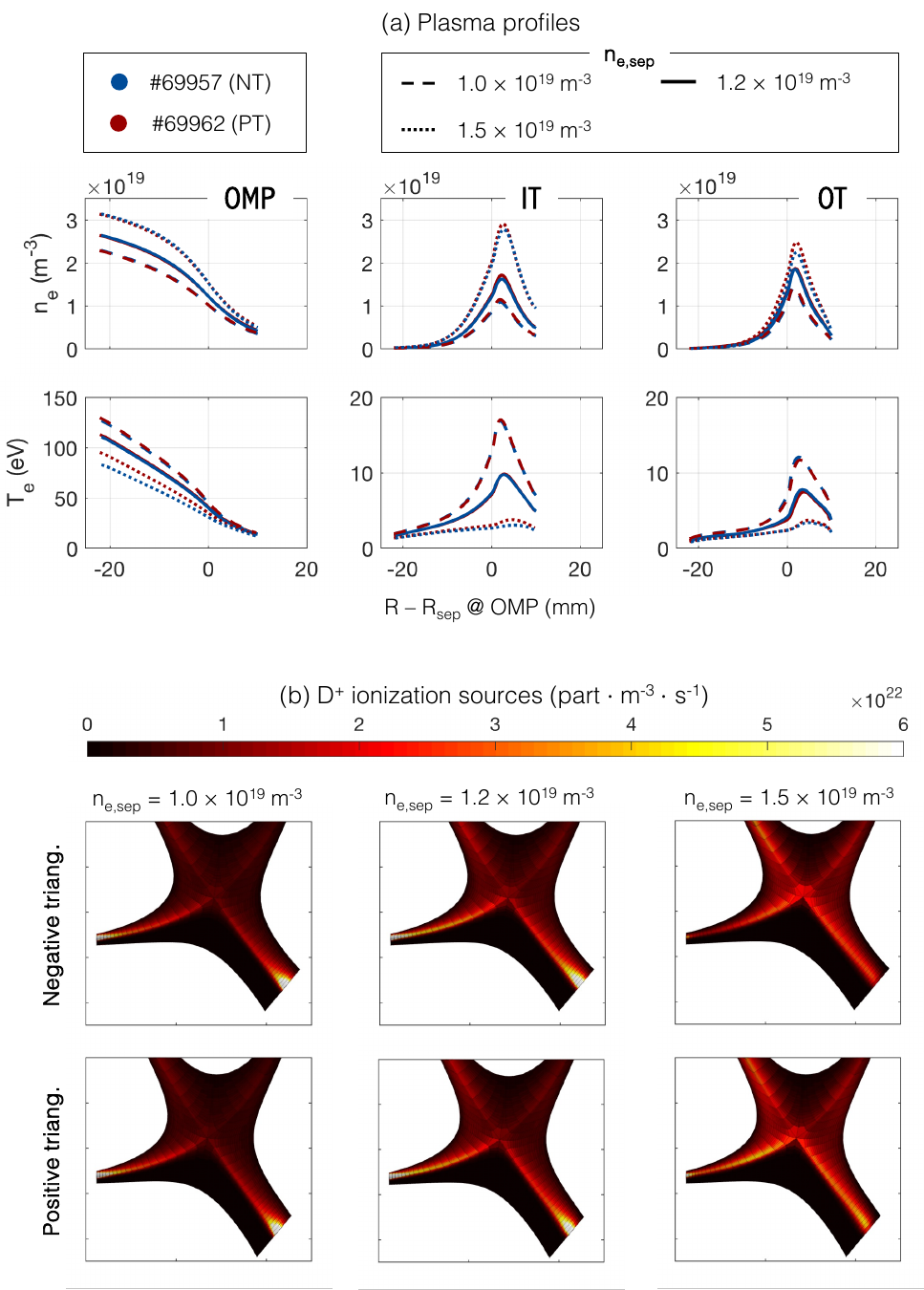}
  \caption{(a) Comparison of electron density and temperature profiles at the OMP and targets provided by SOLPS-ITER simulations performed with identical input parameters for the NT and PT discharges under investigation, for increasing values of electron density at the separatrix ($n_{e,\text{sep}}$). (b) Distribution of ionization sources for \ce{D+} in the divertor region in particles$\cdot$m$^{-3} \cdot$s$^{-1}$ in the same simulations of NT (top row) and PT (bottom row) scenarios.}
  \label{fig:comp_base}
\end{figure}

In Figure \ref{fig:comp_base}(a), the electron density and temperature profiles at the OMP and at the targets, obtained for three different values of separatrix electron density. The profiles are identical both upstream and at the divertor. Similarly, the distribution of ionization sources for \ce{D+} in the divertor region, Figure \ref{fig:comp_base}(b), is similar between the pairs of discharges corresponding to the same \( n_{e,\text{sep}} \), exhibiting a common evolution toward outer divertor detachment in both cases. As the density increases, the OT tends to get cooler and the ionization front progressively moves from the divertor region toward the X-point, following the same trend in both topologies. In the highest-density case, however, the source distribution appears to indicate a less pronounced detachment in the PT configuration—opposite to what is observed experimentally.

\begin{figure} 
  \centering
  \includegraphics[width=\textwidth]{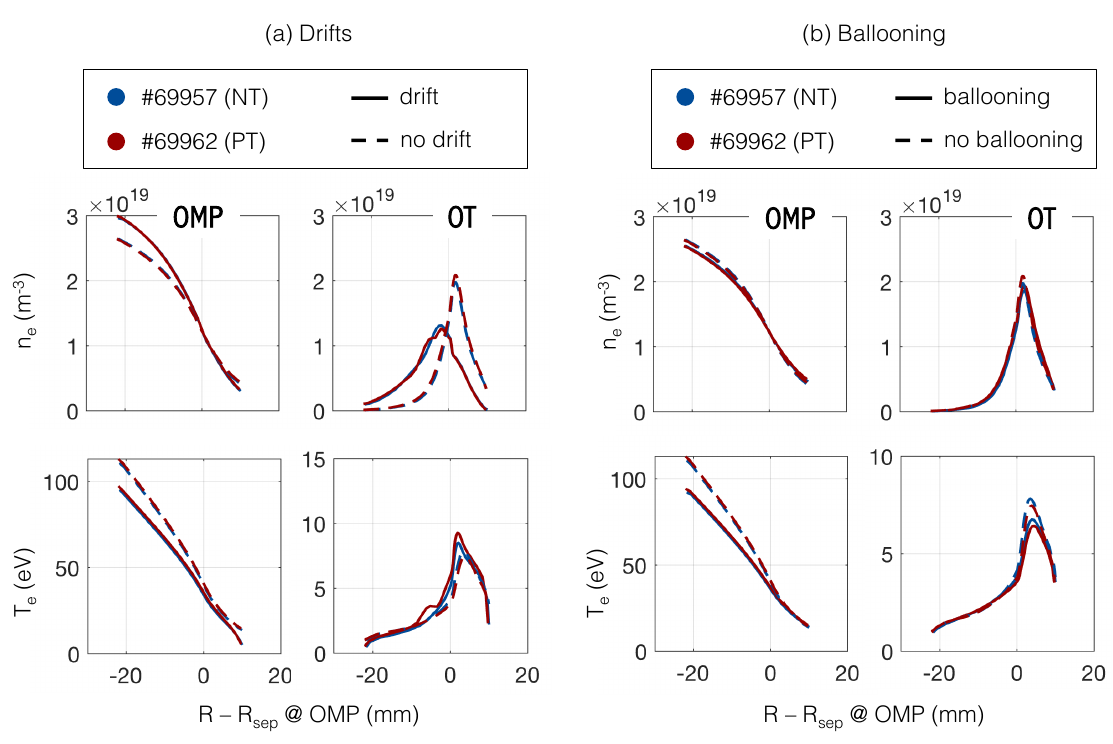}
  \caption{Electron density and temperature profiles at the OMP and OT from SOLPS-ITER simulations performed at fixed input parameters with (a) drifts and (b) ballooning effects enabled for the two scenarios under investigation, compared with the baseline case shown in Figure \ref{fig:comp_base}(a).}
  \label{fig:ball_drift}
\end{figure}

Given the absence of significant differences in the baseline case described, the simulations were extended to incorporate effects associated with drifts. Their inclusion alters the plasma solution both at the OMP—leading to steeper profiles for the same transport coefficients—and at the target, where it affects both the position and the height of the density peak, Figure \ref{fig:ball_drift}(a).

Despite the modification following the activation of drifts, plasma profiles remain indistinguishable between PT and NT. This suggests that the drifts currently implemented in SOLPS-ITER - those primarily associated with the toroidal magnetic field - are not sufficient to account for the experimentally observed differences. A more complete treatment including drifts associated with the poloidal field may be required to fully assess their role.



Finally, a comparison between the plasma profiles is performed by introducing the same poloidal dependence of the anomalous transport coefficients in both scenarios. This feature, available in SOLPS-ITER, is intended to effectively capture the impact of ballooning instabilities.

It is well established that, in magnetically confined plasmas, MHD ballooning instabilities can develop in regions with steep pressure gradients and relatively weak magnetic fields, typically located at the outer edge of the plasma, thereby contributing to confinement degradation.

The anomalous transport coefficients in the cell identified by the indices $(ix, iy)$ are rescaled by a factor that depends on the local magnetic field strength, according to the following expression:

\begin{equation}
\alpha(ix, iy) = a \cdot \left| \frac{B\_{\text{ref}}}{B(ix, iy)} \right|^c
\end{equation}

where $a$ and $c$ are arbitrary constants. This formulation enhances transport where the magnetic field is weaker, consistent with the expected behavior of ballooning-driven transport. When this shaping is activated, using $a = 1$ and $c = 2$, a slight deviation emerges with respect to the simulations without shaping. However, this deviation affects both the PT and NT cases in a similar manner, leading to no appreciable difference between their respective profiles, as shown in Figure \ref{fig:ball_drift}(b).

Based on the results discussed in this subsection, simulations conducted with identical inputs for the PT and NT scenarios revealed no significant differences, indicating that magnetic geometry alone cannot account for the experimentally observed discrepancies.

\subsection{Sensitivity scan of transport coefficients}
\label{scan}

The previous results suggest an intrinsic difference in the transport regime, which is investigated in this section by presenting the results of a parametric scan of the cross-field anomalous transport coefficients for the two discharges, performed without drifts activated\footnote{Note that, given the well-established evidence that simulated plasma profiles in PT and NT are nearly identical for the same input parameters, the sensitivity scans reported here have been performed on the mesh constructed using the magnetic equilibrium of NT.}. This analysis aims to guide us towards optimizing the choice of the input parameters for each scenario and interpreting their physical origin.

This approach is similar to that previously adopted by Muscente et al. \cite{muscente_analysis_2023}. In that study, \( n_{e,\text{sep}} \) was left free to vary during the scan of transport coefficients, potentially leading to a superposition of effects. In the present work, we vary the transport profiles while keeping \( n_{e,\text{sep}} = 1.2 \times 10^{19} \,\text{m}^{-3} \) fixed for both PT and NT cases by applying a feedback condition on the gas puff, as done in Reference \cite{Tonello_phd} for He plasmas in ASDEX Upgrade.

\begin{figure} 
  \centering
  \includegraphics[width=\textwidth]{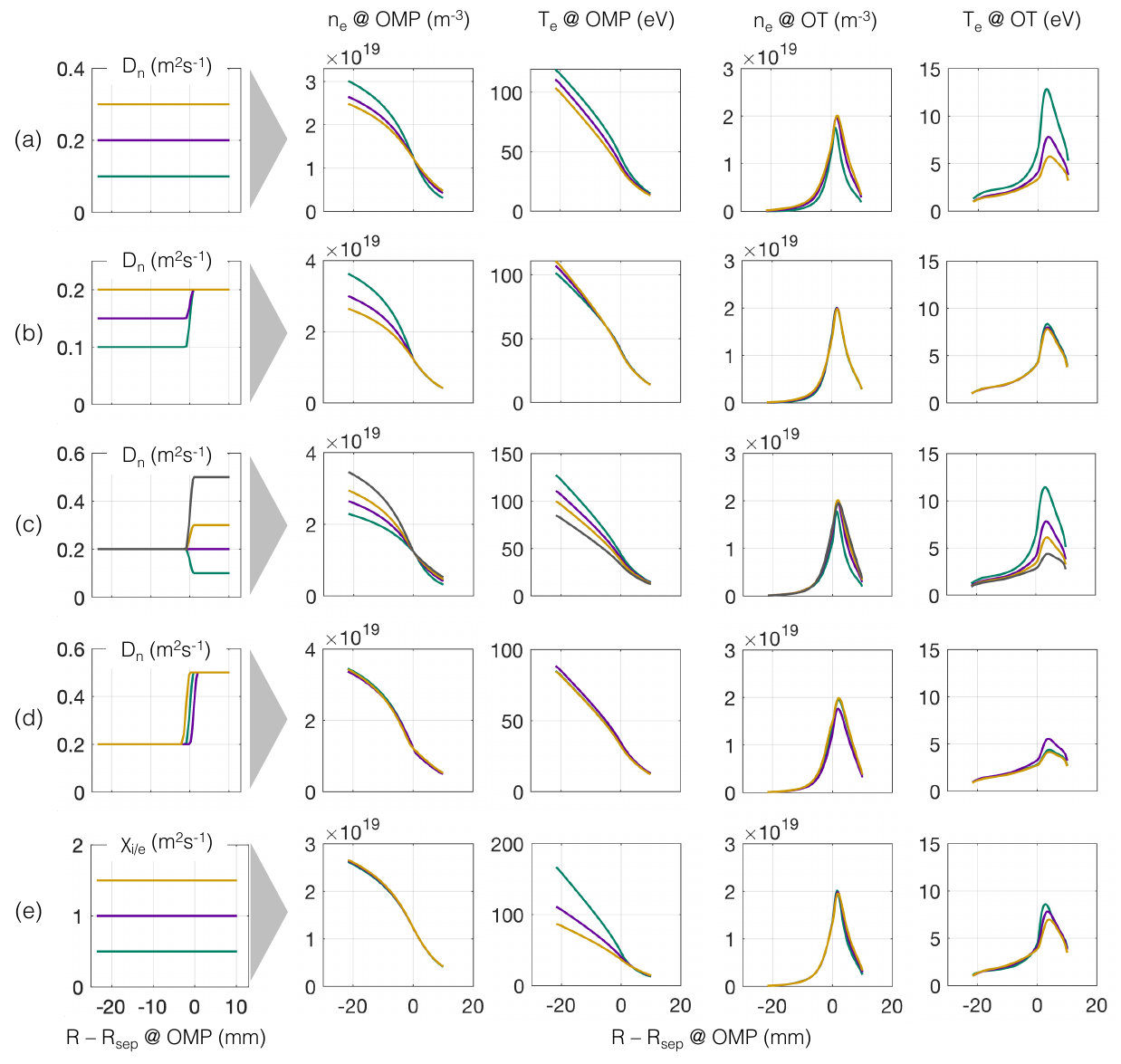}
  \caption{Results of the parametric sensitivity scan of the radial profiles of the particle diffusivity coefficient \(D_n\) and the ion/electron thermal diffusivity \(\chi_{i,e}\), conducted for the negative triangularity baseline simulation (analogous results are obtained for positive triangularity). Each row displays, from left to right, the radial profiles of the investigated transport coefficient, the radial profiles of electron density and temperature at the OMP, and the radial profiles of electron density and temperature at the OT. All input parameters other than the one being scanned are kept fixed at the standard values: \(D_n = 0.2 \, \text{m}^2\text{s}^{-1}\), \(\chi_{i,e} = 1.0 \, \text{m}^2\text{s}^{-1}\), \(n_{e,\text{sep}} = 1.2 \cdot 10^{19} \, \text{m}^{-3}\).}
  \label{fig:parametric_scan}
\end{figure}

Figures \ref{fig:parametric_scan}(a) and \ref{fig:parametric_scan}(e) show the profiles resulting from the scan of particle and heat diffusivities, taken as radially uniform. As $D_n$ increases, the upstream electron density profile tends to flatten the gradients, while the electron temperature decreases. At the OT, an increase in $D_n$ leads to a modest rise in density alongside a noticeable reduction in electron temperature. Conversely, an increase in the thermal diffusivity for electrons and ions leads to a reduction in electron temperature and its upstream gradient, while only a slight corresponding decrease is observed at the targets.

Thus, as suggested by Muscente's work, step-like diffusivity profiles were tested, with the aim of distinguishing the transport in the core periphery from that in the SOL. It is precisely in the core periphery and across the separatrix that numerical studies using turbulence codes indicate suppressed transport for NT \cite{Lim2023}, consistent with its improved confinement properties. Given the dominant effect of particle diffusivity on the profiles, the focus is placed on it while keeping thermal diffusivity constant and radially uniform from now on.


Figures \ref{fig:parametric_scan}(b) and \ref{fig:parametric_scan}(c) show the resulting profiles from a scan of the particle diffusivity values conducted separately in the regions respectively inside and outside the separatrix. In the first case, increasing values of $D_n$ lead to a flattening of the electron density profiles exclusively in the upstream region inside the separatrix, while no significant effects are observed in the SOL and at the targets.

In the second case, as $D_n$ increases in the SOL, a similar effect to that observed for an overall variation of $D_n$ at the divertor is recorded, with electron temperature increasing as $D_n$ decreases. The upstream electron density profiles exhibit a decreasing trend both inside the separatrix (as already observed for uniform $D_n$) and outside the separatrix (in contrast to the behavior observed for uniform $D_n$) as $D_n$ increases. A radial shift in the position of the diffusivity step does not significantly affect the plasma profiles (Figure \ref{fig:parametric_scan}(d)). Repeating the same scans in the presence of drifts yields qualitatively similar results.

\subsection{Optimization of transport coefficients and discussion}

\begin{figure} 
  \centering
  \includegraphics[width=\textwidth]{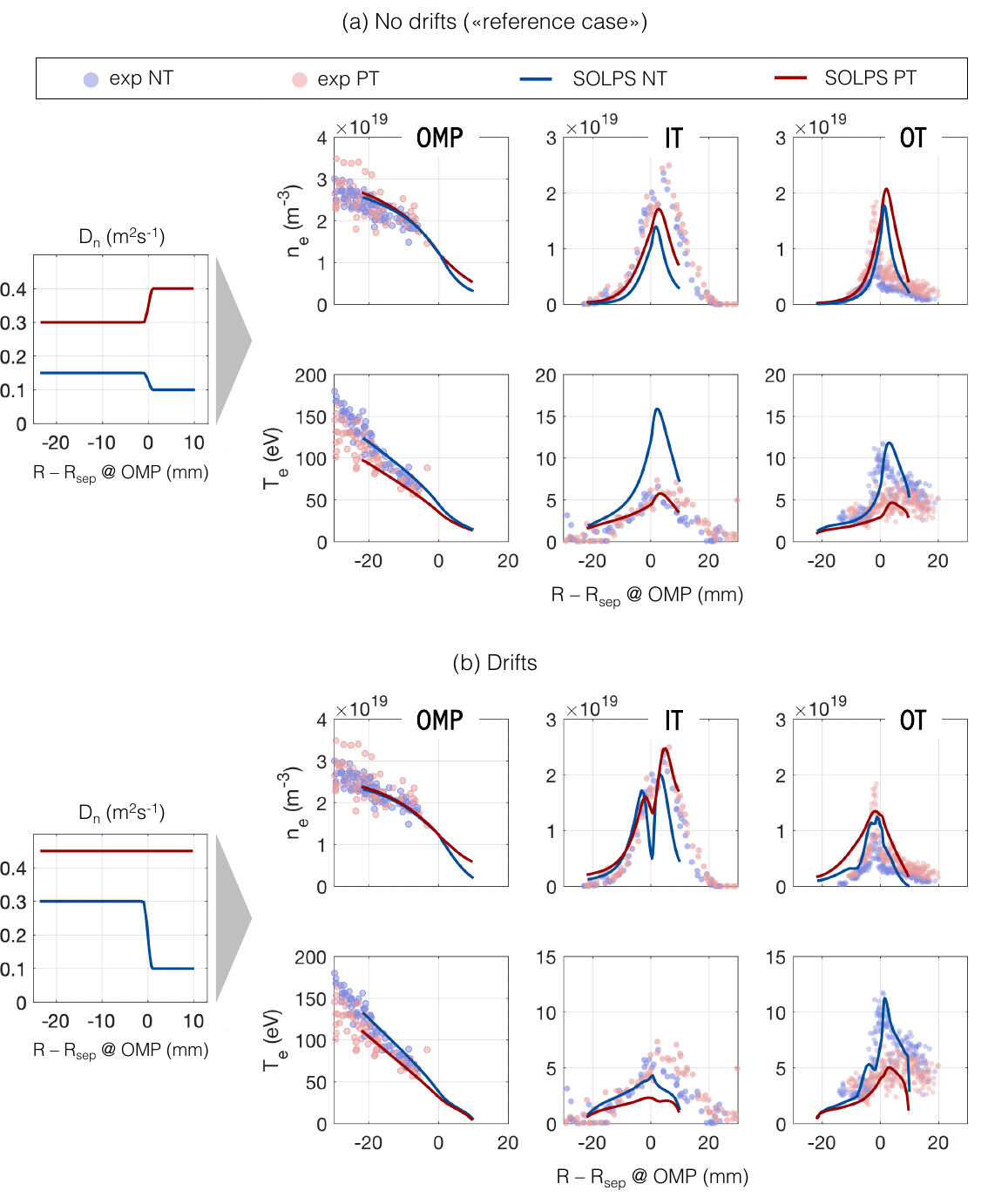}
  \caption{Comparison between electron density and temperature profiles from SOLPS-ITER simulations using optimized particle diffusivity coefficients (reported on the left side) and corresponding experimental data: (a) without drifts and (b) with drifts enabled.}
  \label{fig:optimization}
\end{figure}

Based on the parametric scan presented in Section \ref{scan}, two pairs of optimal radial particle diffusivity profiles were identified—one for simulations without drifts (hereafter referred to as the \textit{reference simulations}) and one including drifts. These profiles were selected according to the criterion that the simulations should accurately and simultaneously reproduce the experimental radial profiles of electron density and temperature at the OMP, as well as the electron temperature at the outer target, for both PT and NT cases (Figure \ref{fig:optimization}).

The identified profiles for transport coefficients are shown in Figure \ref{fig:optimization}(a), with those for PT being consistently higher than those for NT, both in the core region and in the SOL. This result is consistent with the findings of Muscente et al. using the SOLEDGE2D-Eirene code \cite{muscente_analysis_2023} and, more generally, with the well-established experimental observation of suppressed transport in NT compared to PT. Moreover, an increased diffusivity coefficients is required following the activation of drifts to accurately reproduce the radial slope of upstream electron density profiles (Figure \ref{fig:optimization}(b)). Note that within both pairs of simulations, the same values of radially uniform electron and ion heat diffusivity were maintained for PT and NT. However, to improve agreement with the data in the presence of drifts, the magnitude of thermal diffusivity was reduced from 1.0 to 0.7 m$^2$s$^{-1}$.

The upstream electron temperature is steeper in NT than in PT, which is consistent with the fact that - with the same power flowing from the core into the SOL - a higher particle diffusivity in PT results in a greater radial particle flux, associated with a higher convective radial heat flux. Consequently, for the same poloidal heat flux and anomalous thermal conductivity, a lower radial temperature gradient is sufficient to conduct the remaining power.

The electron temperature profiles at the OT accurately reproduce the experimental data, with \( T_e \) in NT being larger than in PT, which falls below 5 eV, suggesting the onset of a detachment regime, which is not observed in NT.  

Regarding the electron density at the OT, simulations without drifts correctly capture the experimental trend of higher electron density in PT compared to NT. However, the difference between the two cases is underestimated, and the experimental radial position of the density peak is not reproduced. Including drifts accurately reproduces the radial position of the density peak, which appears to be significantly influenced by the \( E \times B \) drift-induced flux. Under forward field conditions, this flux is directed into the PFR at the outer target, determining a corresponding cross-field shift of the peak.

It should be noted that the transport profiles were optimized to match the density and temperature profiles upstream and at the outer target, while allowing the model to self-consistently describe the profiles at the inner target (IT). Experimentally, higher density and lower temperature are observed at the inner divertor compared to the outer divertor, as typically occurs under forward field conditions.  

In this regard, simulations without drifts tend to underestimate the electron density and significantly overestimate the electron temperature in NT at the inner target. Only with the inclusion of drifts—considered the primary drivers of the asymmetry between the inner and outer divertor—does the predicted electron density at the IT quantitatively match experimental values. This also results in an underestimation of electron temperature, which falls below the detachment threshold for both discharges. Nonetheless, the NT simulation results remain consistent with the observation that the inboard divertor tends to detach at a lower core density than the outboard divertor.

\begin{figure} 
  \centering
  \includegraphics[width=0.85\textwidth]{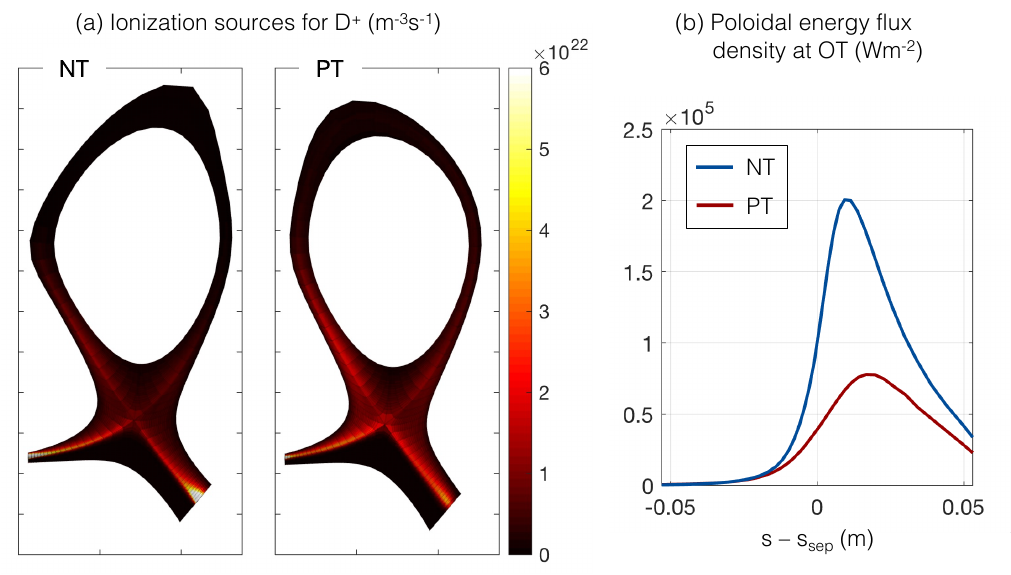}
  \caption{For the reference simulations without drifts of the scenarios under investigation, (a) distribution over the B2.5 computational meshes of the ionization source intensity for the \ce{D+} species, expressed in particles $\cdot$m$^{-3} \cdot$s$^{-1}$; (b) poloidal heat flux density profiles at the outer target, as a function of the distance along the target measured from the separatrix.}
  \label{fig:ioniz_opt_hf}
\end{figure}

Comparing the distribution of \ce{D+} ionization source for the reference simulations, Figure \ref{fig:ioniz_opt_hf}(a), ionization occurs predominantly in the immediate vicinity of the target in NT, defining a typical high-recycling regime. Conversely, in PT, the ionization front tends to extend along the outer divertor leg toward the X-point, another characteristic signature of the onset of the detached regime.

\begin{figure} 
  \centering
  \includegraphics[width=0.8\textwidth]{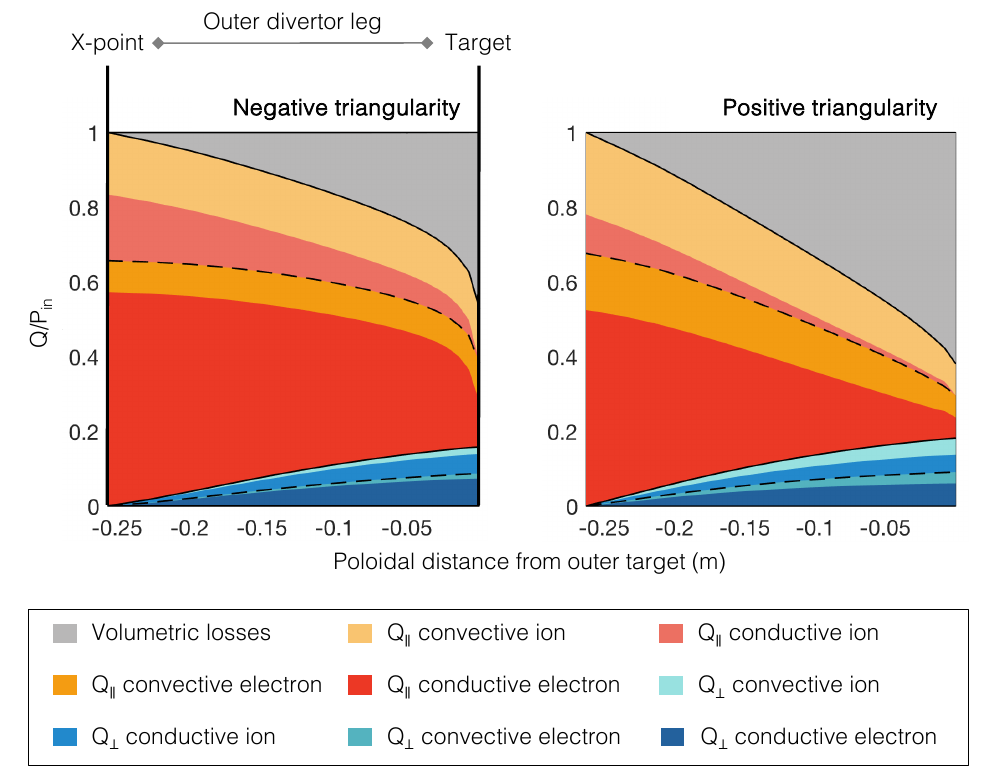}
  \caption{Schematic representation of the energy balance for the reference simulations of the discharges under investigation, between the X-point and the outer divertor. The total power entering from the upstream surface associated with the X-point, denoted as P$_{in}$, is analyzed in terms of its transport mechanisms along the outer divertor leg, highlighting the fraction $Q/P_{in}$ that is transported parallel to the magnetic field lines $Q_{\parallel}$, eventually reaching the divertor plate, the one removed by radial transport and lost at the far SOL boundary of the computational mesh $Q_{\perp}$, and the one lost through volumetric processes. Both parallel and perpendicular transport channels are further decomposed into their conductive and convective components, with separate contributions from ions and electrons.}
  \label{fig:energy_balance}
\end{figure}

Figure \ref{fig:energy_balance} schematically represents the various contributions to the overall energy balance and transport between the X-point and the outer divertor for the reference simulations. In PT, a smaller fraction of the heat flux entering at the X-point\footnote{Note that the plot shows the contributions to the energy balance and transport between the X-point and the outer divertor, normalized to the power entering through the radial surface crossing the X-point. Although the input power through the core boundary is the same in both cases, different dissipation occurs already upstream of the X-point—for instance, due to different convective losses at the far SOL boundary of the mesh, resulting from the different transport coefficients applied. This leads to different absolute values of the power entering the X-point. However, the normalization allows for a direct comparison of the energy transport and dissipation mechanisms along the outer divertor leg, highlighting the impact of access (or lack thereof) to the detachment regime.} reaches the divertor target due to more efficient thermal dissipation via volumetric processes occurring along the outer divertor leg, including atomic processes, radiation, and plasma-neutral interactions. A stronger radial convective contribution is also observed in PT, consistent with the higher anomalous particle diffusivity imposed in this configuration.

From the comparison of the radial profiles of heat flux density at the outer divertor (Figure \ref{fig:ioniz_opt_hf}(b)), in addition to confirming that the heat flux in PT is reduced due to more efficient dissipation, it is evident that PT is associated with a larger power fall-off length, which leads to a broader distribution of power reaching the plate. This result is consistent with experimental observations \cite{Lim2023,faitsch_dependence_2018}.



\subsection{Density scan and neutral pressure analysis}

Having selected transport coefficient profiles that yield a satisfactory agreement with experimental data for both discharges under reference conditions, with and without drifts included, this section presents the results of a separatrix density scan.

The scan is carried out by varying \( n_{e,\text{sep}} \)—the control parameter of the gas puff feedback scheme—within the range \( 0.9 \times 10^{19} \) to \( 1.5 \times 10^{19} \) m\(^{-3} \), while keeping the same transport coefficients fixed. This range was chosen to encompass the experimental conditions observed during the density ramp of the discharge.

Figure \ref{fig:density_ramp_validation} displays the electron density and temperature profiles obtained from the drift-including simulations at values of \( n_{e,\text{sep}} \) of (a) \( 0.9 \times 10^{19} \,\mathrm{m}^{-3} \) (“low density”) and (b) \( 1.5 \times 10^{19} \,\mathrm{m}^{-3} \) (“high density”), which are, respectively, lower and higher than the reference value of \( 1.2 \times 10^{19} \,\mathrm{m}^{-3} \) used in the baseline simulations.

\begin{figure} 
  \centering
  \includegraphics[width=0.85\textwidth]{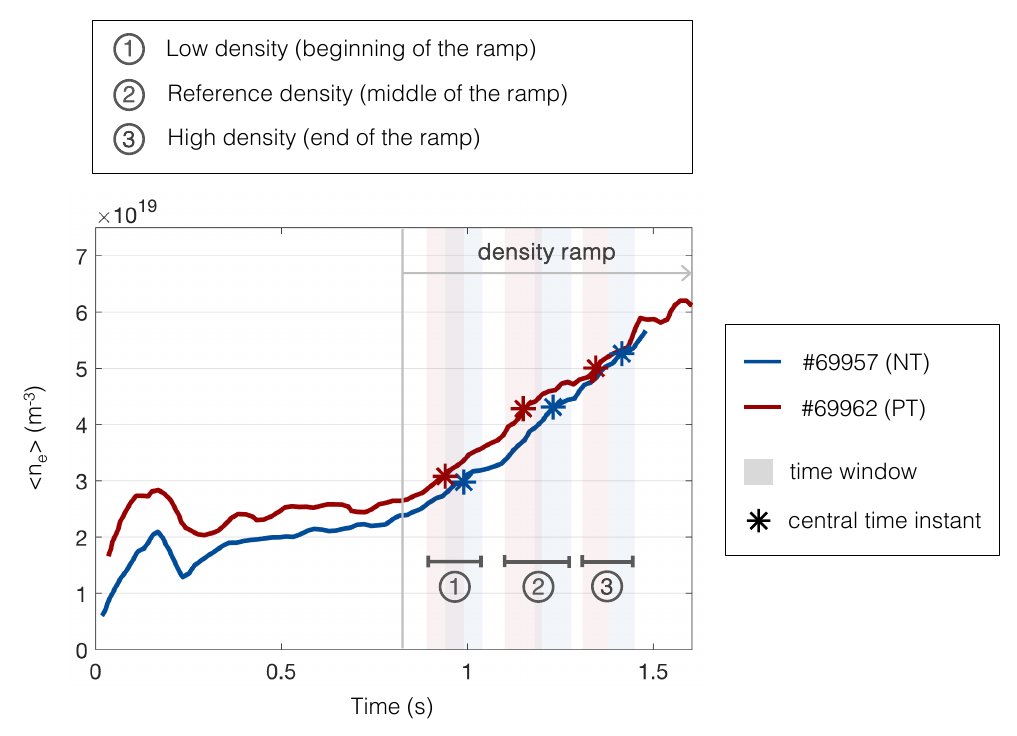}
  \caption{Temporal evolution of the line-averaged electron density during the discharges considered for the comparison between positive and negative triangularity. The discharges feature a stationary phase followed by a density ramp. Time windows and representative time points are selected at the beginning, middle, and end of the ramp, ensuring comparable line-averaged density values between the two configurations. These points correspond to \textit{reference}, \textit{low} and \textit{high} density conditions used for comparison with SOLPS-ITER simulations in Figures \ref{fig:optimization} and \ref{fig:density_ramp_validation}.}
  \label{fig:ramp_instants}
\end{figure}

\begin{figure} 
  \centering
  \includegraphics[width=\textwidth]{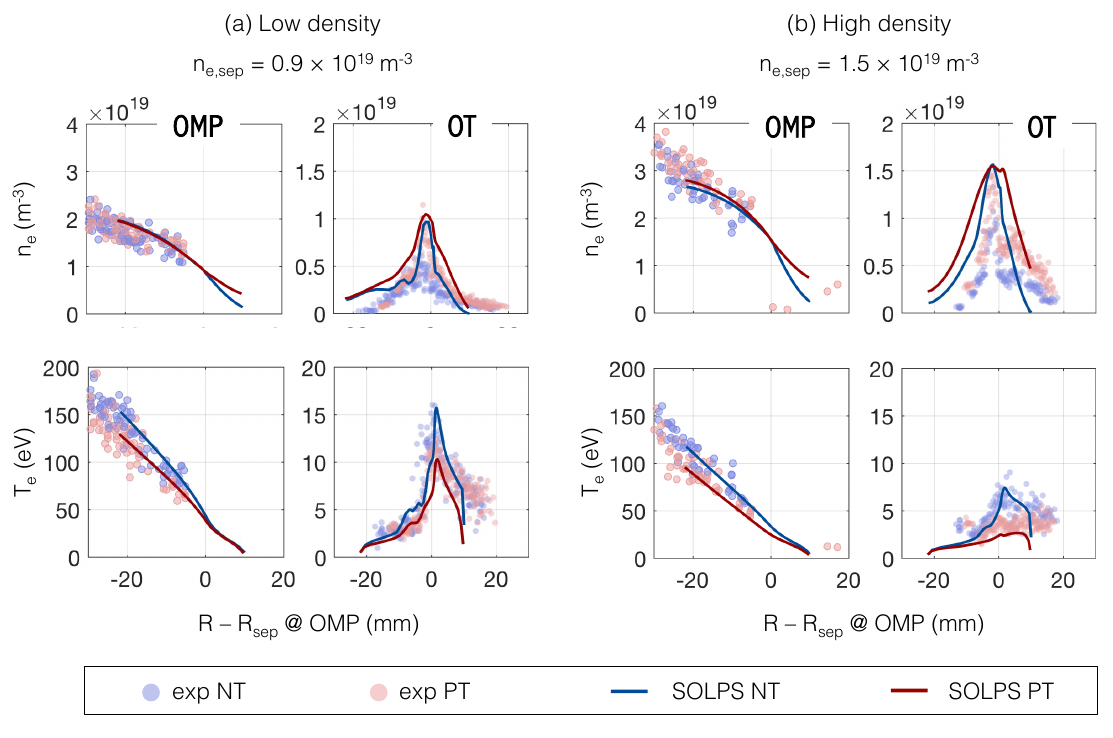}
  \caption{Comparison between electron density and temperature profiles at the OMP and OT from SOLPS-ITER simulations, performed by varying the separatrix electron density, and experimental profiles obtained under (a) low- and (b) high-density plasma conditions, corresponding to points 1 and 3 along the experimental density ramp as shown in Figure \ref{fig:ramp_instants}.}
  \label{fig:density_ramp_validation}
\end{figure}

The numerical profiles are compared with experimental data extracted within specific time windows during the density ramp, identified in Figure \ref{fig:ramp_instants}. These time points were selected to represent the beginning and end of the ramp, while ensuring that the average density was comparable between the two magnetic configurations. For comparison, the experimental conditions corresponding to low, reference, and high density approximately match the first three columns shown in Figure 13 of Reference \cite{Fvrier2024}.

The agreement between the simulations and the experimental data remains good even under conditions different from the reference ones. This suggests a consistent transport regime throughout the density scan.

The density scan is also used to extract the evolution of divertor neutral pressure in the two scenarios, with the aim of assessing whether SOLPS-ITER reproduces the experimental trend and provides a physical interpretation of the observed behavior.

To extract a single neutral pressure value from the simulations for comparison with the experimental measurement, an assumption about neutral transport from the chamber to the baratron gauge is required. This is achieved through a synthetic diagnostic based on the analysis of kinetic neutral fluxes provided by EIRENE, following the approach previously adopted in Reference \cite{Tonello2024}.

Since, for the reasons discussed in Section \ref{reference_exp_frame}, the experimental radial density profiles upstream are limited to the region inside the separatrix, the comparison between the experimental and synthetic neutral pressure has been performed by mapping it as a function of the electron density evaluated at the magnetic coordinate \(\rho = 0.95\)\footnote{The normalized poloidal magnetic flux coordinate is defined as $\rho = \sqrt{\frac{\psi - \psi_0}{\psi_{\text{sep}} - \psi_0}} $, where \(\psi\), \(\psi_0\), and \(\psi_{\text{sep}}\) are the poloidal magnetic fluxes at the radial location of interest, at the magnetic axis, and at the separatrix, respectively. By construction, $\rho = 0$ at the magnetic axis and $\rho = 1$ at the separatrix. The position $\rho = 0.95$ lies just inside the separatrix, in the edge region of the plasma core.}.

\begin{figure} 
  \centering
  \includegraphics[width=\textwidth]{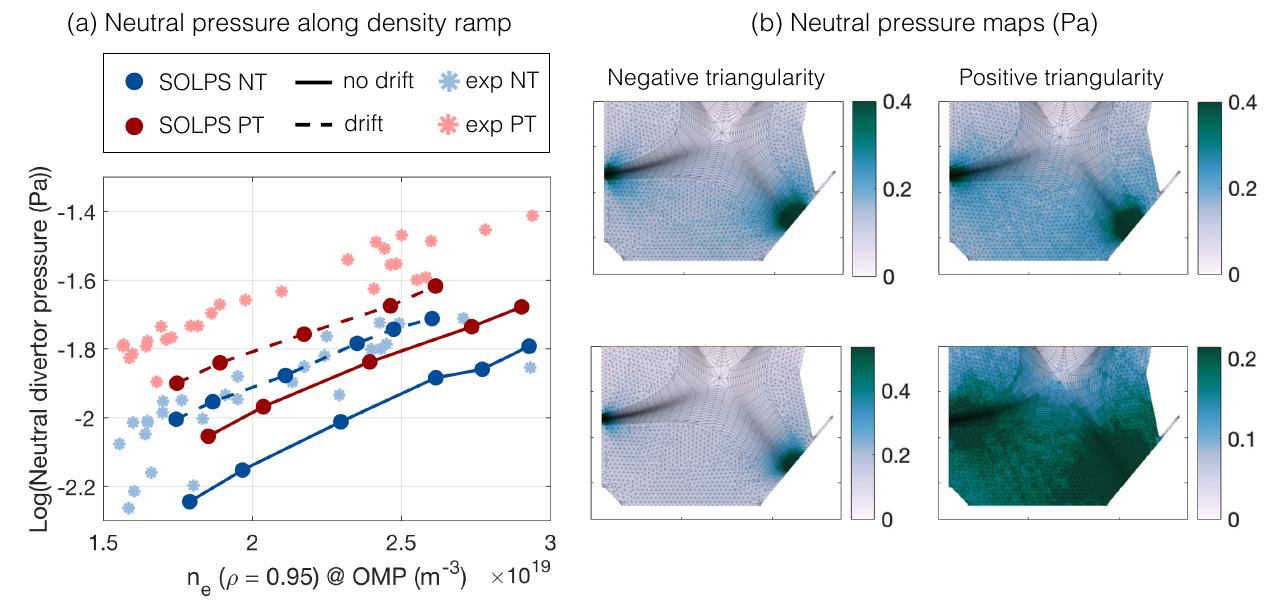}
  \caption{(a) Neutral pressure at the divertor as a function of electron density at $\rho=0.95$ during the density ramp for the two discharges under investigation. Comparison between experimental values obtained with a baratron gauge and SOLPS-ITER results processed through synthetic diagnostics. (b) Neutral pressure distribution in the divertor region extracted from the two reference simulations presented in Figure \ref{fig:optimization}(a). In the top row, both colorbars are capped at the same upper value. In the bottom row, the colorbars for each simulation are individually capped at the 95th percentile value of the pressure distribution within the divertor region.}
  \label{fig:pressure_comp}
\end{figure}

The synthetic neutral pressure values accurately reproduce the experimental trend (Figure \ref{fig:pressure_comp}(a)). The synthetic neutral pressure increases with $n_e \left (\rho = 0.95 \right)$, and is systematically higher in PT compared to NT, both in simulations without and with drifts. While both sets of simulations correctly capture these trends, the simulations with drifts are closer to the absolute experimental values\footnote{In this representation, the delay between the measurement of upstream electron density profiles by Thomson scattering and the measurement of neutral pressure at the divertor is assumed to be negligible.}.

Figure \ref{fig:pressure_comp}(b) shows the 2D map of neutral pressure over the computational domain of EIRENE, including the atomic contributions from \ce{D} and \ce{C} species as well as the molecular contribution from \ce{D2}, for the reference simulations. These correspond to the reference condition of \( n_{e,\text{sep}} = 1.2 \times 10^{19} \) particles \(\cdot\) m\(^{-3}\) and no drifts for both NT and PT scenarios. The first row displays the maps with the color scale limited to the same value, clearly showing a qualitatively similar distribution in both cases, with peaks at the two targets where neutral recycling occurs, and approximately uniform distribution throughout the lower portion of the PFR. However, the absolute values of neutral pressure throughout the entire PFR, which the baratron duct samples, are overall lower in NT compared to PT, which justifies the results yielded by the synthetic baratron diagnostic.

In the second row, the same pressure map is shown with the color scale limited to the 95th percentile of the neutral pressure values provided by EIRENE for the cells in the divertor region, thus favoring a comparison focused not on absolute pressure values, but rather on their statistical distribution. The neutral pressure in the immediate vicinity of the outer target is higher in NT than in PT. However, the pressure values in the PFR are overall closer to the neutral pressure at the targets in PT, outlining a situation in which the neutral pressure throughout the entire divertor region tends to be relatively uniform. Conversely, in NT, the neutral pressure values in the PFR exhibit a significant drop compared to the corresponding values at the targets.

\begin{figure} 
  \centering
  \includegraphics[width=\textwidth]{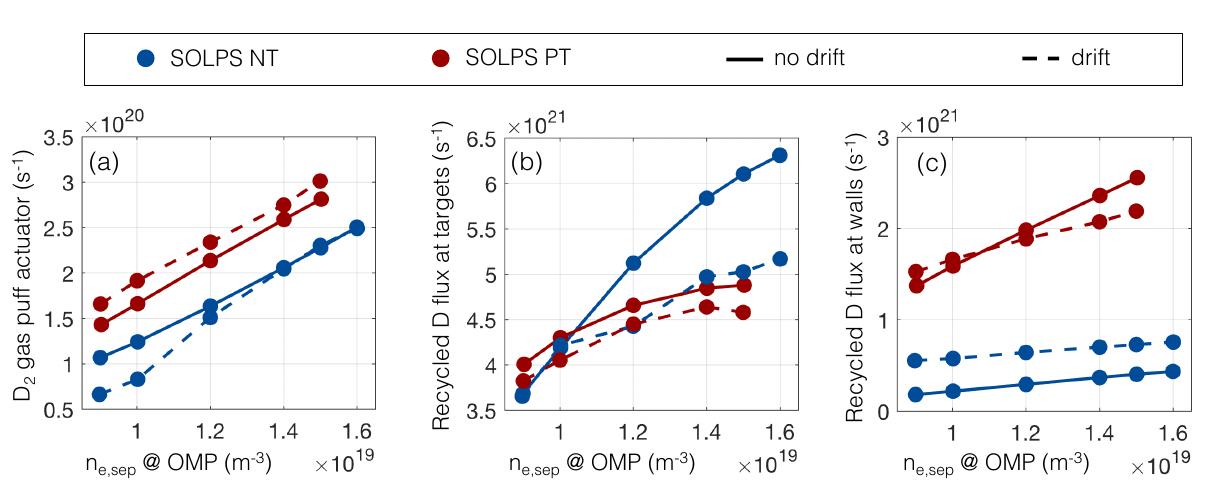}
  \caption{(a) Gas puff actuator strength, (b) neutral deuterium recycling fluxes, in both atomic and molecular form, at the targets and (c) at the walls (i.e., at the boundaries of the fluid mesh), as provided by SOLPS-ITER simulations performed without and with drifts along a density ramp, shown as a function of the electron density imposed at the upstream separatrix and expressed in \ce{D} equivalent atoms/s.}
  \label{fig:recycling}
\end{figure}

This distinct distribution can be interpreted in light of the balance between recycling and ionization processes. The analysis of neutral deuterium recycling fluxes along the density scan (Figure \ref{fig:recycling}(b)) reveals that, in the NT case, recycling not only tends to be higher than in the PT configuration—consistent with the larger neutral pressures observed at the targets—but also increases significantly at the divertor targets as the upstream separatrix electron density rises, coherently with a high-recycling regime. Conversely, the PT simulations indicate a saturation of the ion flux to the targets and the associated recycling, which is characteristic of detachment conditions.

On the other hand, the \ce{D+} ionization source remains more localized near the target in the NT case. As a consequence, in PT, the ionization mean free path is likely longer, enabling neutrals produced by recycling at the target to be transported more effectively into the private flux region (PFR).

This interpretation is supported by the analysis of the \ce{D2} gas puff rate required to achieve the prescribed \(n_{e,\text{sep}}\) in the simulations, which is consistently higher in PT than in NT (Figure \ref{fig:recycling}(a)). While one can assume that the gas puff has a limited impact on the neutral pressure in the divertor region—being more than one order of magnitude lower than the recycling flux at the targets—the fact that it is higher in PT suggests that the detached ionization source is less effective at refueling the upstream plasma, thereby requiring an increased gas puff. In other words, a greater amount of neutrals is needed in PT to maintain the same electron density at the separatrix upstream.


Finally, the flux of recycled \ce{D} at the far SOL and PFR boundaries of the B2.5 computational mesh (Figure \ref{fig:recycling}(c)) is higher in PT, which is consistent with the greater radial ion flux resulting from the higher imposed transport coefficients, given a comparable plasma density gradient.

\section{Conclusions}
\label{conclusions}

This paper presented the comparison and analysis of two TCV discharges featuring matching magnetic geometry at the divertor and opposite upper triangularity, using the edge plasma code SOLPS-ITER. The aim was to investigate the physical mechanisms underlying the experimental observation of a more challenging access to the detachment regime of the outer divertor leg and, consequently, reduced target cooling in NT compared to PT under identical upstream conditions \cite{Fvrier2024}.

To determine whether the sole difference in magnetic configuration is sufficient to account for the experimentally observed discrepancies, simulations of the two discharges were performed using computational meshes constructed based on their respective magnetic equilibria, with identical input parameters, in particular the transport coefficients used to describe anomalous transport in the SOL. These simulations did not reveal significant differences between the two cases, even when including drift or providing anomalous diffusivities with a poloidal shaping, yielding essentially identical plasma profiles both upstream and at the target.

Having established that the sole difference in magnetic geometry is not sufficient to account for the experimental evidence, the initial assumption of identical transport regimes in the two cases was relaxed. A parametric scan of anomalous particle diffusivity values was then conducted, adopting a step-like profile to decouple transport in the core from that in the SOL, while maintaining the working assumption of a fixed upstream separatrix electron density. Based on this scan, a better match of the numerical plasma profiles relative to experimental data was achieved by assuming lower particle diffusivity in NT compared to PT, both in the core and in the SOL, with and without drifts activated. This result is consistent with findings from other codes \cite{muscente_analysis_2023} and with the experimental observation of improved plasma confinement in NT.

The comparison of these simulations with experimental data reveals a reduction in electron temperature at the OT in PT below 5 eV under the same upstream conditions as NT, indicating access to the detachment regime. This is further confirmed by the displacement of the ionization front from the target toward the X-point. Additionally, the heat flux to the divertor in NT is not only higher, consistent with reduced volumetric dissipation, but also associated with a smaller fall-off length \( \lambda_q \).

Finally, the simulation of a separatrix density scan, carried out with fixed transport coefficients, consistently reveals higher neutral pressure values at the divertor in PT compared to NT, in agreement with experimental observations and consistent with the divertor detachment regime accessed in the PT configuration.

\section*{Acknowledgements}

The authors would like to acknowledge Xavier Bonnin for his constant support and Antonello Zito for the fruitful interactions. This work has been carried out within the framework of the EUROfusion Consortium (WPTE), partially funded by the European Union via the Euratom Research and Training Programme (Grant Agreement No 101052200 — EUROfusion). The authors wish to acknowledge Benoit Labit and Olivier Sauter for their roles as Task Force Leader and Scientific Coordinator, respectively, of EUROfusion WPTE RT02. The Swiss contribution to this work has been funded by the Swiss State Secretariat for Education, Research and Innovation (SERI). Views and opinions expressed are however those of the author(s) only and do not necessarily reflect those of the European Union, the European Commission or SERI. Neither the European Union nor the European Commission nor SERI can be held responsible for them.

\newpage

\appendix

\section{Technical and operational considerations on the use of drifts in TCV SOLPS-ITER simulations}

\label{appendix}

This appendix contains a series of observations regarding the use of drifts in the SOLPS-ITER simulations presented in this work. As previously mentioned, to the best of our knowledge, this study represents the first application of the so-called partial flux surface averaging convergence acceleration technique to the TCV tokamak. The activation of this technique requires modifying specific switches in the input files \texttt{b2mn.dat} and \texttt{b2.numerics.parameters}, in addition to implementing appropriate drift-compatible boundary conditions \cite{Kaveeva2018}. This approach allows simulations with an active feedback scheme and impurity modeling to reach convergence within approximately two days when running the MPI-based code on 16 cores.

\begin{figure} 
  \centering
  \includegraphics[width=\textwidth]{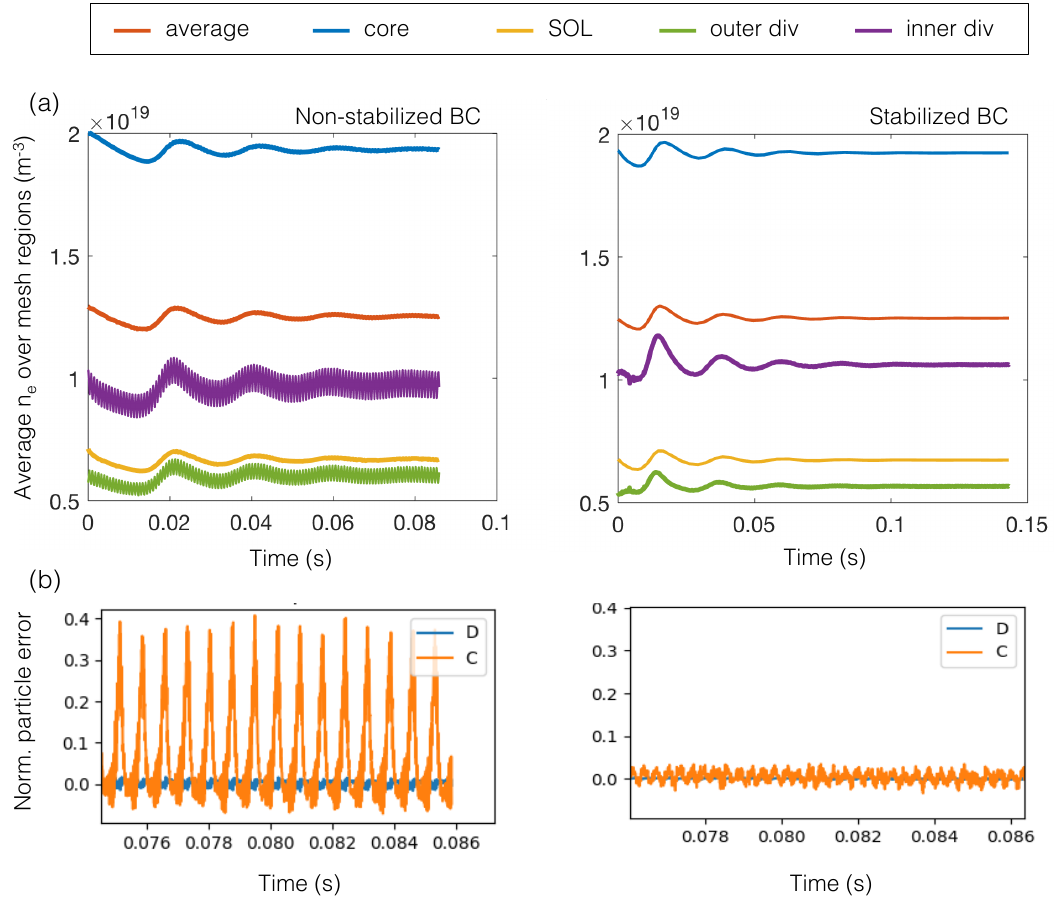}
  \caption{(a) Time evolution, during a negative triangularity simulation, of the average electron density in different regions of the B2.5 computational mesh. Left: case with boundary conditions without stabilization scheme. Right: case with stabilized boundary conditions. (b) Corresponding time evolution of the normalized particle error.}
  \label{fig:nereg}
\end{figure}

A first observation concerns the evolution of the average electron density in the different regions of the B2.5 domain during the simulation (\texttt{nereg}), shown in the Figure \ref{fig:nereg}(a). It is observed that the density values tend to stabilize around asymptotic values in the different regions of the B2.5 computational mesh. However, at the targets, high-frequency oscillations persist.  

This introduces a significant dependence of the magnitude of the electron density profiles at the targets on the specific iteration at which the simulation is interrupted once convergence is reached. Nevertheless, it should be noted that the other profiles under examination—namely, the upstream density and temperature, as well as the temperature at the targets—are unaffected by this variability associated with the electron density at the target. Based on the particle balance analysis, these oscillations appear to originate from the balance of carbon impurities (Figure \ref{fig:nereg}(b)).

To mitigate these transient oscillations, some available stabilization techniques for drift-compatible boundary conditions on electron and ion energy at the mesh boundaries corresponding to the targets were tested. These techniques introduce a stabilizing term to the source associated with the boundary conditions and can be activated by specifying the appropriate flags in the \texttt{b2mn.dat} input file.

\begin{verbatim}
'b2stbc_stab_coeff_sheath_te' '50'
'b2stbc_stab_coeff_sheath_ti' '50'
\end{verbatim}

In addition, an optional stabilization mechanism is enabled to assist the linearization of the leakage boundary condition (\texttt{BCCON 10}), which is applied to all species at the computational mesh boundaries corresponding to the private flux region (PFR) and the far scrape-off layer (SOL). This option is activated by specifying a positive parameter via the following setting in the \texttt{b2.boundary.parameters} file. This approach proved effective, successfully suppressing the previously observed oscillations (Figure \ref{fig:nereg}(a)).

\begin{verbatim}
CONPAR(,,2) = 1
\end{verbatim}

\begin{figure} 
  \centering
  \includegraphics[width=0.9\textwidth]{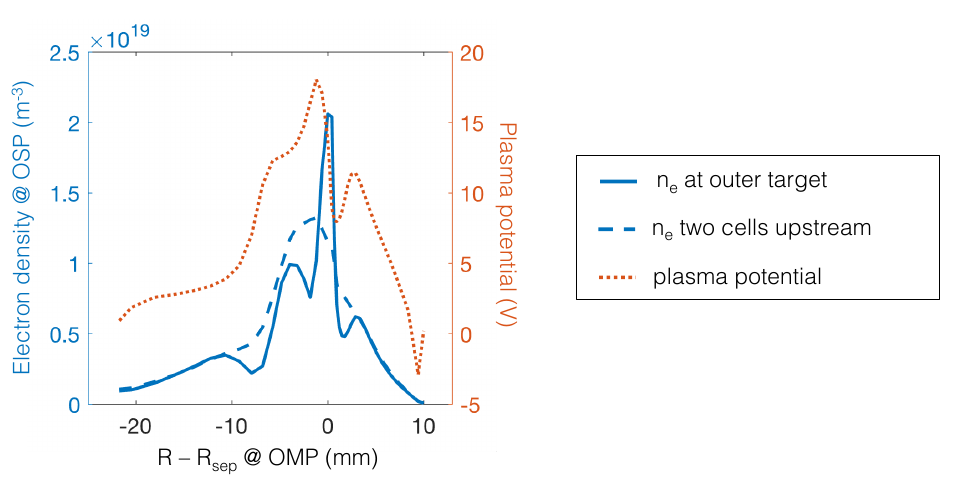}
  \caption{Comparison between electron density profiles at the outer target extracted from a negative triangularity simulation with drifts and a stabilization scheme enabled at the target and a few cells upstream. The electron density profiles are shown together with the corresponding plasma potential.}
  \label{fig:peaks_drift}
\end{figure}

Despite suppressing transient oscillations, simulations with drifts consistently exhibit multiple peaks in the electron density radial profiles at both targets. Double peaks in electron density at the divertors in SOLPS-ITER simulations with drifts have already been observed in TCV simulations and in other cases within specific density regimes \cite{Colandrea2024, Jia2022}. It is consistently observed that the largest peak aligns with the maximum gradient of the plasma potential and the resulting electric field, suggesting a possible role of the poloidal velocity pattern associated with the \( E \times B \) drift (Figure \ref{fig:peaks_drift}).

However, it is observed that if the radial electron density profiles at the targets are extracted not from the last - poloidally - series of cells (corresponding to the guard cells at the target boundary) but rather from a couple of cells upstream of the target, the characteristic multiple peaks disappear, and the profiles appear much smoother (Figure \ref{fig:peaks_drift}). 

This suggests a potentially non-physical nature of these peaks—affecting only the density profiles—and indicates that they may instead arise from numerical issues related to the type and implementation of boundary conditions. For this reason, the electron density profiles at the outer target shown in the figures of this paper for drift-enabled simulations are extracted two cells upstream.

\newpage

\section*{Reference}
\bibliographystyle{iopart-num}
\bibliography{Bibliography.bib}  

\end{document}